\documentclass[11pt]{article}
\usepackage{epsfig}
\begin{document}
\setlength{\baselineskip}{0.30in}
\newcommand{\simgt}{\,\rlap{\lower 3.5 pt \hbox{$\mathchar \sim$}} \raise
1pt \hbox {$>$}\,}
\newcommand{\simlt}{\,\rlap{\lower 3.5 pt \hbox{$\mathchar \sim$}} \raise
1pt \hbox {$<$}\,}
\newcommand{\nc}{\newcommand}
\newcommand{\beq}{\begin{equation}}
\newcommand{\eeq}{\end{equation}}
\newcommand{\be}{\begin{eqnarray}}
\newcommand{\ee}{\end{eqnarray}}
\newcommand{\bi}{\bibitem}
\newcommand{\dlnk}{\partial_{{\rm ln} k}}
\def\app#1#2#3{ Astroparticle Phys. {\bf #1} (#2) #3}
\def\Aa#1#2#3{ Astron. Astrophys. {\bf #1} (#2) #3}
\def\aj#1#2#3{ Astron. J. {\bf #1} (#2) #3}
\def\apj#1#2#3{ Astrophys. J. {\bf #1} (#2) #3}
\def\araa#1#2#3{ Annu.~Rev.~Astron.~Astrophys. {\bf #1} (#2) #3}
\def\ass#1#2#3{ Astrophys. Space Sci. {\bf #1} (#2) #3}
\def\aspr#1#2#3{ Astrophys. Space Phys. Rev. {\bf #1} (#2) #3}
\def\casp#1#2#3{ Comments Astrophys. Space Phys. {\bf #1} (#2) #3}
\def\ib#1#2#3{ ibid. {\bf #1} (#2) #3}
\def\mn#1#2#3{ Mon. Not. R. Astron. Soc. {\bf #1} (#2) #3}
\def\nps#1#2#3{ Nucl. Phys. B (Proc. Suppl.) {\bf #1} (#2) #3}
\def\np#1#2#3{ Nucl. Phys. {\bf #1} (#2) #3}
\def\nat#1#2#3{ Nature {\bf #1} (#2) #3}
\def\pl#1#2#3{ Phys. Lett. {\bf #1} (#2) #3}
\def\prd#1#2#3{ Phys. Rev. D {\bf #1} (#2) #3}
\def\prep#1#2#3{ Phys. Rep. {\bf #1} (#2) #3}
\def\prl#1#2#3{ Phys. Rev. Lett. {\bf #1} (#2) #3}
\def\rmp#1#2#3{ Rev. Mod. Phys. {\bf #1} (#2) #3}
\def\zp#1#2#3{ Z. Phys. {\bf #1} (#2) #3}
\def\sjnp#1#2#3{ Sov. J. Nucl. Phys. {\bf #1} (#2) #3}
\def\jetp#1#2#3{ Sov. Phys. JETP {\bf #1} (#2) #3}
\def\jetpl#1#2#3{ JETP Lett. {\bf #1} (#2) #3}
\def\jhep#1#2#3{ J. High Energy Phys. {\bf #1} (#2) #3}
\def\ppnp#1#2#3{ Prog. Part. Nucl. Phys. {\bf #1} (#2) #3}
\def\ptp#1#2#3{ Prog. Theor. Phys. {\bf #1} (#2) #3}
\def\rpp#1#2#3{ Rep. Prog. Phys. {\bf #1} (#2) #3}
\def\yf#1#2#3{ Yad. Fiz. {\bf #1} (#2) #3}

\makeatletter
\def\alt{\mathrel{\mathpalette\vereq<}}
\def\vereq#1#2{\lower3pt\vbox{\baselineskip1.5pt \lineskip1.5pt
\ialign{$\m@th#1\hfill##\hfil$\crcr#2\crcr\sim\crcr}}}
\def\agt{\mathrel{\mathpalette\vereq>}}
\def\lsim{\raise0.3ex\hbox{$\;<$\kern-0.75em\raise-1.1ex\hbox{$\sim\;$}}}
\def\gsim{\raise0.3ex\hbox{$\;>$\kern-0.75em\raise-1.1ex\hbox{$\sim\;$}}}
\makeatother
\begin{center}
{\bf \Large
Probing the power spectrum bend with recent CMB data
}
\bigskip
\\{\bf S.~Hannestad
\footnote{e-mail: {\tt steen@nordita.dk}}\\
{\small {\it{NORDITA, Blegdamsvej 17, DK-2100 Copenhagen, Denmark
\\
}}}}
{\bf S.H. Hansen \footnote{e-mail: {\tt hansen@astro.ox.ac.uk}} \\
{\small {\it{NAPL, University of Oxford, Keble road, OX1 3RH, Oxford, UK
\\
}}}}
{\bf F.L.~Villante  \footnote{e-mail: {\tt villante@fe.infn.it}}
\\
{\small
{\it{Dipartimento di Fisica and INFN, Via del Paradiso 12,
44100 Ferrara, Italy\\
}}}}
\end{center}

\begin{abstract}
We constrain the spectrum of primordial curvature perturbations 
${\cal P}(k)$ by using the new data on the Cosmic Microwave Background (CMB) 
from the Boomerang and MAXIMA experiments. 
Our study is based on slow-roll inflationary models,
and we consider 
the possibility of a running spectral index. Specifically,
we expand the power spectrum ${\cal P}(k)$ to second order in $\ln k$, thus
allowing the power spectrum to ``bend'' in $k$-space. We
show that allowing the power spectrum to bend erases the ability of
the present data to measure the tensor to scalar perturbation
ratio. Moreover, if the primordial baryon density 
$\Omega_b h^2$ is as low as found from Big Bang nucleosynthesis (BBN), 
the data favor a negative bending of the power spectrum,
corresponding to a bump-like feature in the power spectrum 
around a scale of $k=0.004\, {\rm Mpc}^{-1}$.
\end{abstract}    

PACS: 98.70.Vc, 98.80.Cq

\section{Introduction}
Inflation is generally believed to set the stage for the evolution of
the universe, in particular providing the initial conditions for
structure formation and cosmic microwave background (CMB)
anisotropies.  From a given inflationary model one can calculate the
power spectrum of primordial curvature perturbations, 
${\cal P}(k)$, which is a function of the
wavenumber $k$.  This power spectrum can then be Taylor expanded about
some wavenumber $k_0$ 
and truncated after a few terms~\cite{Lidsey:1997np}
\be
{\rm ln} {\cal P} (k) = {\rm ln} {\cal P} (k_0) +
(n - 1)\,
{\rm ln} \frac{k}{k_0} +\left. \frac{1}{2} \frac{d\, n}{d \, {\rm ln}k}
\right|_{k_0} \, {\rm ln} ^2 \frac{k}{k_0} + \cdots
\label{power}
\ee
where the first term corresponds to a scale invariant
Harrison-Zel'dovich spectrum,
the second is the power-law approximation, and the third term is the
running of the spectral index, which will be our main concern below.
In order to provide the nearly scale-invariant perturbations, which
seem to be observed, one probably has to consider a slow-roll (SR)
model for the later stage of inflation. The properties of SR models
are well-known, and to set notation we recollect the basic features
(see Ref.~\cite{Lyth:1999xn} for a review and list of references). In
SR models one demands that the first few derivatives of the inflaton
potential should be small. Traditionally this is expressed with the 3
SR parameters $(\epsilon, \eta, \xi^2)$, which roughly correspond to
the first, second and third derivative of the potential. With these SR
parameters one can express the scalar spectral index, $n(k) - 1 \equiv 
d\ln {\cal P}/d{\rm ln}k$, and tensor spectral index, $n_T(k)$,
and their derivatives~\cite{Liddle:1992wi,Kosowsky:1995aa}.  
We will here adopt a slightly different notation, 
and instead of the set $(\epsilon,
\eta, \xi^2)$ we will use the 3 parameters $(n, r, \dlnk)$, 
where $n \equiv d\ln {\cal P}/d{\rm ln}k|_{k=k_0}+1$ 
is the scalar spectral index at the scale $k_0$, the parameter 
$r$ is the tensor to scalar perturbation ratio at the 
quadrupole, and $\dlnk \equiv dn/d{\rm
ln}k|_{k=k_0}$.  The reason is simply that these 3 variables are
closer related to what is being observed. With these 3 parameters one
automatically expresses~\cite{Lyth:1999xn,Kosowsky:1995aa}
 the tensor spectral index and its
derivative in $k_0$ as
\be
n_T = - \frac{r}{6.8} \, \, \, \, \, {\rm and}  \, \, \, \, \, 
\frac{d\, n_T}{d \, {\rm ln} k} = \frac{r}{6.8} \left[ \left( n-1 \right) 
+ \frac{r}{6.8} \right] ~.
\label{consist}
\ee
The factor 6.8 in the above equation is actually model-dependent
and should be calculated for each given model,
in particular for different $\Omega_\Lambda$~\cite{Turner:1996ge}, but
for simplicity we use the fixed value 6.8.

The anisotropies, which are integrals over all the wavenumbers, pick
up the major contribution from $k \approx l H_0/2$, and hence one
finds~\cite{Kosowsky:1995aa}
\be
\frac{C_l\left[ n(k)\right]}{C_l\left[ n(k_0)\right]} 
\approx \left( \frac{l}{l_0} \right)^{\frac{1}{2}{\rm ln} (l/l_0)\,\dlnk}~,
\ee
where $l_0\approx 2 k_0/H_0$, which means that for a running spectral
index, $\dlnk \neq 0$, the power spectrum will be {\em bent} (up or
down depending on the sign of $\dlnk$), besides the normal {\em tilt}
which arises for $n \neq 1$. 
In many SR models $\dlnk$ is expected to 
be very small since it is second
order in the small parameters, however, there are very interesting
models where this need not be the case~\cite{stewart,Kinney:1998dv},
and $\dlnk$ may assume values big enough to be observable~(see
e.g. Refs.~\cite{Copeland:1997mn,Covi:1999mb}).

Let us briefly mention where in our parameter space the different
inflationary models lie. Traditionally~\cite{Dodelson:1997hr} one
divides the simplest inflationary models into the following groups:
small field $(\alpha<-1)$, large field $(-1<\alpha<1)$, and hybrid
models $(\alpha>1)$, where $\alpha$ is defined through the SR
parameters (see Appendix A for notation)
\be
\eta \equiv \alpha \, \epsilon~\, \, \, \, \, {\rm where} \, \, \, \, \, 
\epsilon = \frac{r}{13.6} \, \, ,  \, \, 
\eta = \frac{n-1}{2} + 0.15 \, r~. 
\label{alpha_eq}
\ee
\begin{figure}[htb]
\begin{center}
\epsfig{file=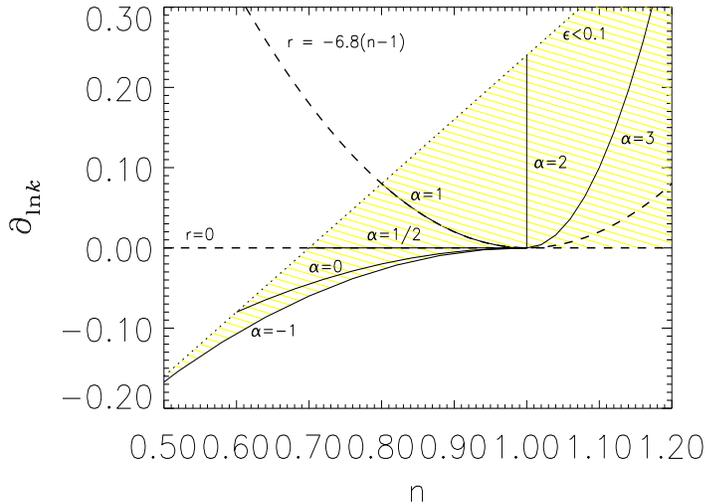,height=8cm,width=10cm}
\end{center}
\caption{The various slow-roll models presented in $(n,\dlnk)$ space. The
dotted line is the conservative limit, $\epsilon < 0.1$. The two
dashed lines are the two attractors.}
\label{fig1}
\end{figure}
If we for simplicity consider models where the third derivative of the
potential is zero ($\xi^2 =0 $), then the different inflationary
models are placed in the $(n,\dlnk)$-space as shown in Fig.~1, where
we have plotted various curves for different $\alpha$. Here we have
allowed only the conservative limit $\epsilon<0.1$ (showed by the
dotted line), and assumed that we only have to expand the expressions
to leading order.  Naturally the graph moves up and down by inclusion
of the third derivative of the potential, $\xi^2 \neq 0$.  By changing
the pivot scale, $k_0$, around which the power spectrum is expanded,
one can also move the graph sideways (see discussion in
section~\ref{dataanalysis}), and the $n$ in Fig.~1 should therefore be
thought of as the $n$ where the expansion scale has been chosen near
the center of the probed scales.

It is interesting that there seems to be attractors in the SR
parameter space~\cite{hoffman}. The two attractors found
in~\cite{hoffman} can be expressed as $r=0$ and $r=-6.8(n-1)$, and the
behavior of these solutions in the plane $(n,\dlnk)$ is shown in
Fig.~1 (dashed lines). As one can clearly see, according to these
results $\dlnk$ should be positive (or zero).  Again, including the
third derivative of the potential will also allow for a negative
$\dlnk$. From Fig.~2 one sees how these two attractors (dashed lines)
are on the borders between the different inflationary models.
\begin{figure}[htb]
\begin{center}
\epsfig{file=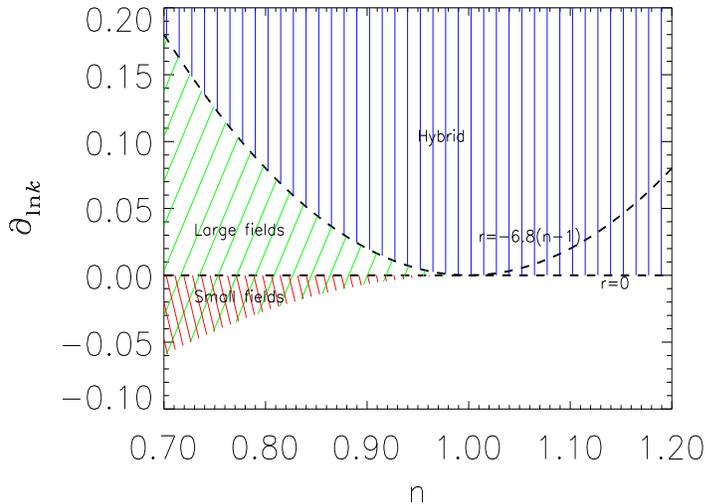,height=8cm,width=10cm}
\end{center}
\caption{The various slow-roll models in $(n,\dlnk)$ space. The hybrid models 
correspond to $\alpha>1$, large fields means $-1 < \alpha < 1$, and
small fields means $\alpha < -1$. The dashed lines are the two
attractors.}
\label{fig2}
\end{figure}

We will in the present paper discuss the ability of the present day
data to provide information on the 3 parameters, $(n, r, \dlnk)$.


\section{The data}

Anisotropies in the CMB were detected for the first time in 1992 by
the COBE satellite \cite{COBE}. Recently, however, our knowledge of
the temperature perturbations has increased dramatically with the
results from the two balloon-borne experiments Boomerang~\cite{boom}
and MAXIMA~\cite{max}.  One of the main conclusions from these
experiments is the confirmation of the position of the first acoustic
peak at $l \sim 200$, which strongly indicates a flat universe,
$\Omega_{\rm tot} = 1$. This seems to be a confirmation of the
inflationary paradigm, since $\Omega_{\rm tot} = 1$ is a rather
general prediction of the simplest inflationary models.  Very
interestingly, the data also suggest that the second acoustic peak in
the power spectrum is much less pronounced than predicted in flat
models with baryon density compatible with BBN
\cite{jaffe,tegmark}. This could be an indication of new physics
and accordingly a large number of papers on this subject have appeared
since the release of the Boomerang and MAXIMA data. One possibility is
that the primordial spectrum of fluctuations produced by inflation is
not described by a smooth power-law, but rather that it has bumps 
and wiggles
\cite{griffiths,barriga,wang}. In the next
section we will discuss this possibility further in light of our
numerical results.

The data from Boomerang and MAXIMA was recently analyzed in
Ref.~\cite{Kinney:2000nc}, where a search in $(n,r)$ space was
performed.  In that work~\cite{Kinney:2000nc} the assumption was made
that $\dlnk=0$, which means that for each set of $(n,r)$ one must
carefully adjust the third derivative of the potential to make
$\dlnk=0$. This is perfectly possible, but it is more natural to allow
$\dlnk$ to vary, as we will do below.  The effect of running of the
{\em tensor} spectral index is very small, but for consistency we
include it as described in Eq.~(\ref{consist}).

\section{Data analysis}
\label{dataanalysis}

In order to to investigate how the new CMB data constrain the
parameter space $(n,r,\dlnk)$ we have performed a likelihood
analysis of the data sets from COBE \cite{COBE}, Boomerang \cite{boom}
and MAXIMA \cite{max}.  The likelihood function to be calculated is
\begin{equation}
{\cal L} \propto A \exp \left(-\sum_i \frac{(C_{l,i}(\theta)-
C_{l,obs,i})^2}
{\sigma^2(C_{l,i})} \right),
\label{eq:like}
\end{equation}
where $i$ refers to a specific data point and $\theta$ is a vector of
cosmological parameters for the given model
\begin{equation}
\theta = \{\Omega_m,\Omega_\Lambda,\Omega_b,H_0,\tau,Q,n_s,r,\dlnk,\ldots \}~.
\end{equation}
In the present case we have calculated the likelihood function for the
following parameter space: $\Omega_m$, the matter density, $\Omega_b$,
the baryon density, $H_0$, the Hubble parameter, as well as the
inflationary parameter space $n,r,\dlnk$.  
We have assumed that the
universe is flat $\Omega_\Lambda = 1-\Omega_m$, and that reionization
is not important, $\tau=0$.
We also treat the overall normalization, $Q$, as a free parameter.
The experimental groups quote estimated calibration errors for the
experiments. These should also be accounted for, and we do this by
allowing the data points to be shifted up or down by this amount:
10\% for Boomerang \cite{boom} and 4\% for MAXIMA \cite{max}.
It should also be noted that the data points are to some extent
correllated. The likelihood function, Eq.~(\ref{eq:like}), is based
on the assumption that errors are uncorrellated, and therefore a
(quite small) error is introduced into our analysis. 
However, until the full
data sets from the experiments are publicly available, the magnitude 
of the error is difficult to quantify.

Finally, we have chosen the pivot scale in Eq.~(\ref{power}) 
as $k_{0}=0.05\,{\rm Mpc}^{-1}$. This choice is made for convenience, 
since $k_{0}=0.05\,{\rm Mpc}^{-1}$ is the scale at which wavenumbers 
are normalized in the CMBFAST code. Our main results, 
however, do not depend on the specific value of $k_{0}$.

In Fig.~3 we show the allowed region in the inflationary parameter
space for the case corresponding to the calculation of
Ref.~\cite{Kinney:2000nc}, namely $\dlnk=0$.  In the left panel the
allowed region was derived assuming that $\Omega_b h^2 = 0.019$ (BBN
prior
\cite{BBN}), and the
left panel of our Fig.~3 thus corresponds to their Fig.~3.  Our result
is almost identical to theirs, the only difference being that they
have used a slightly larger space of other cosmological parameters, so
their constraints are slightly less restrictive than ours. 
In Fig.~4 of Ref.~\cite{Kinney:2000nc}, $\tau$ was
allowed to vary. Changing the value of $\tau$ does change the likelihood
contours (pushing them to larger values of $n$), 
but only to a small extent. This illustrates that allowing
$\tau$ as a free parameter in our analysis would lead to slightly
larger allowed regions, but would not in any way invalidate our conclusions.

One thing
which should be noticed is that even the best fit point has
$\chi^2/{\rm d.o.f.} = 1.2$, indicating that it is a quite poor
fit. This is not too surprising because it is well known that no good
fit to the new CMB data can be obtained with a low baryon density,
even allowing $r$ to vary \cite{jaffe,tegmark}.

\begin{figure}[hbt]
\begin{center}
\epsfig{file=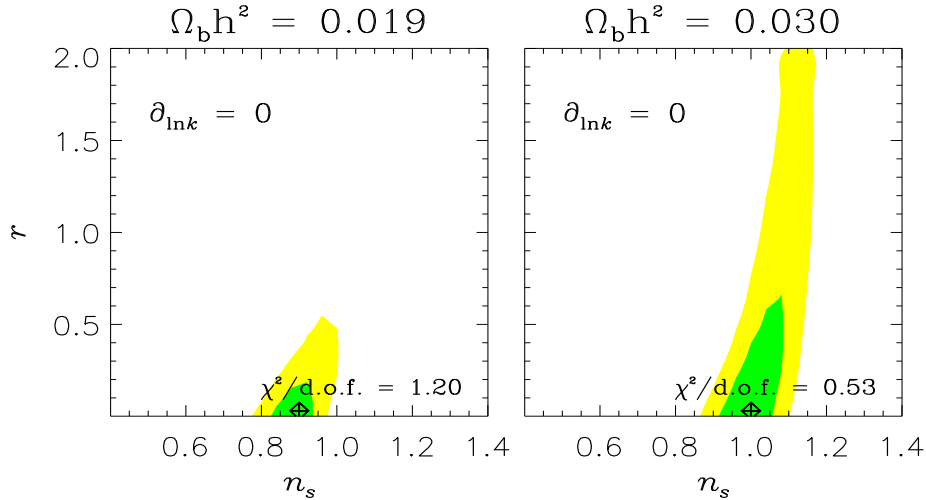,height=7cm,width=11cm}
\end{center}
\vspace*{-0.5cm}
\caption{The allowed region in the $n,r$ parameter space, calculated from
the combined COBE, Boomerang and MAXIMA data.  
The dark shaded (green) regions are $1\sigma$ and the light shaded
(yellow) are $2\sigma$.
The left panel assumes
a BBN prior on $\Omega_b h^2 = 0.019$, whereas the right panel is for
$\Omega_b h^2 = 0.030$, the value which best fits the CMB data.  Note
that the best-fit points, marked by diamonds, are really at $r=0$, but
have been shifted slightly so that they are more visible.}
\label{fig3}
\end{figure}

In the right panel we show the allowed region for the case where
$\Omega_b h^2 = 0.030$, which is the value favored by the CMB
measurements. In this case we find an allowed region very similar to
that found in Fig.~2 of Ref.~\cite{Kinney:2000nc}, although again our
allowed region is slightly smaller than theirs because we use a
smaller parameter space. It should also be noticed that the best fit
point now has $\chi^2/{\rm d.o.f.} = 0.53$, a very good fit.  This
corresponds well to the findings of other likelihood analyses
\cite{jaffe,tegmark}, that the CMB data can be very well fitted in
models with high baryon density.

If the assumption $\dlnk=0$ is relaxed, the results change
substantially. In Fig.~4 we show results for the likelihood analyses
where $\dlnk$ is allowed to vary. Again, the left panels correspond to
$\Omega_b h^2 = 0.019$ and the right panels to $\Omega_b h^2 = 0.030$.
As one can see, we have varied the inflationary parameters
in a wide range, even larger that what allowed by SR approximation.
The reason is that we believe that the main conclusions that we obtain 
are quite general and give indications on the shape of ${\cal P}(k)$ 
which are relevant even outside the context of SR models.

In the upper panels of Fig.~4 we show the same allowed region as in
Fig.~3, when the assumption $\dlnk=0$ is relaxed.  When $\dlnk$ is
allowed to vary, the preferred region in parameter space is shifted
completely.  The best fits are still for a model with no tensor
component (for both values of $\Omega_b h^2$), but now there is no
real constraint on $r$.  The best fit models now have $\chi^2/{\rm
d.o.f.} = 0.54$ and 0.57 respectively, which indicate very good fits.

\begin{figure}[b]
\begin{center}
\epsfig{file=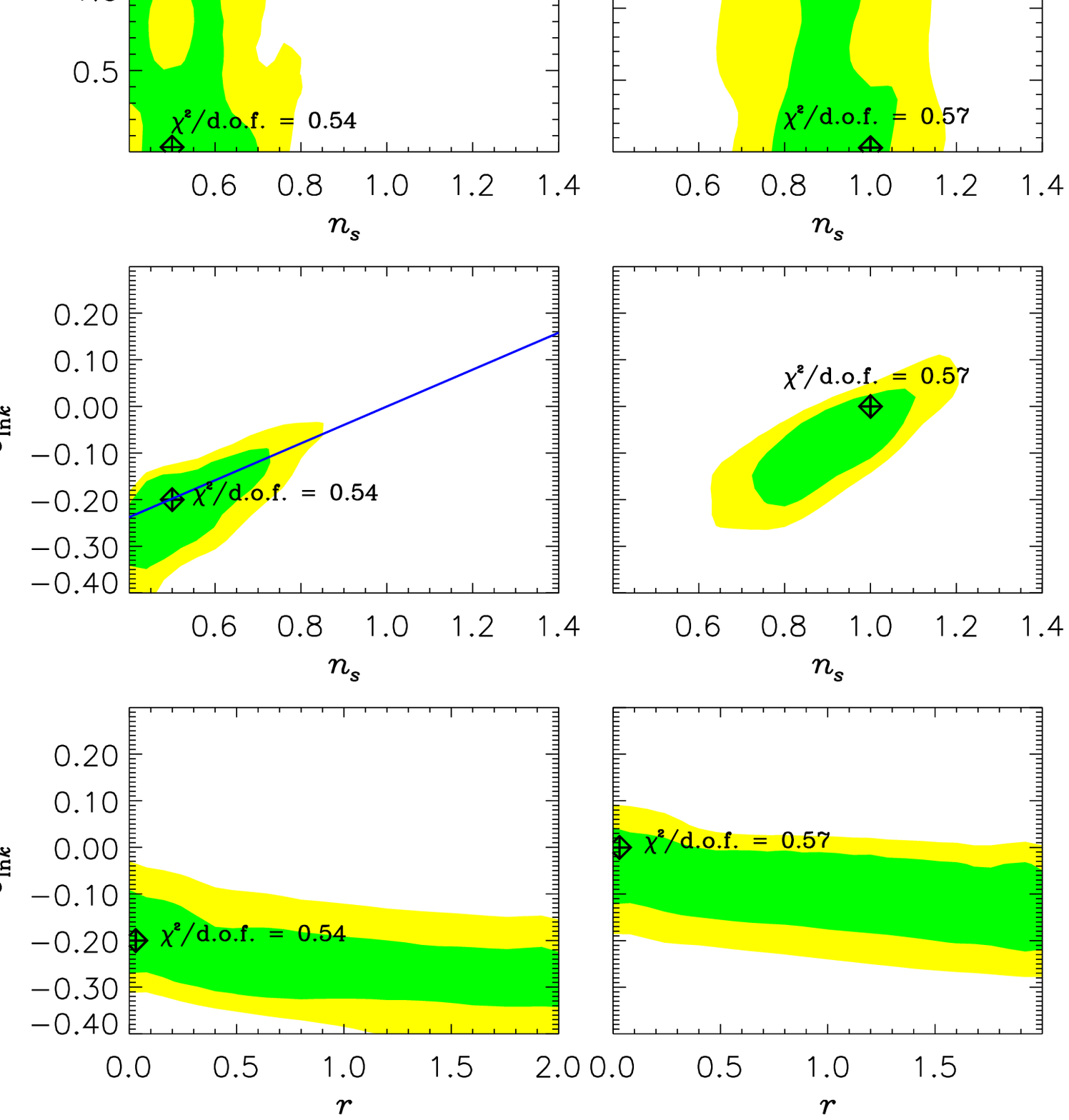,height=14cm,width=10.23cm}
\end{center}
\caption{The allowed region in the $n,r,\dlnk$ parameter space, 
calculated from the combined COBE, Boomerang and MAXIMA data. 
The dark shaded (green) regions are $1\sigma$ and the light shaded
(yellow) are $2\sigma$.
The
left panels assume a BBN prior on $\Omega_b h^2 = 0.019$, whereas the
right panels are for $\Omega_b h^2 = 0.030$, the value which best fits
the CMB data.  Note that the best-fit points, marked by diamonds, are
really at $r=0$, but have been shifted slightly so that they are more
visible.
The solid line in the left $(n,\dlnk)$ panel corresponds to the power
spectrum exhibiting a distinct feature at $k = 0.004 \, {\rm Mpc}^{-1}$
(see Eq.~(8) and the surrounding discussion).}
\label{fig4}
\end{figure}

\clearpage

\begin{figure}[htb]
\begin{center}
\epsfig{file=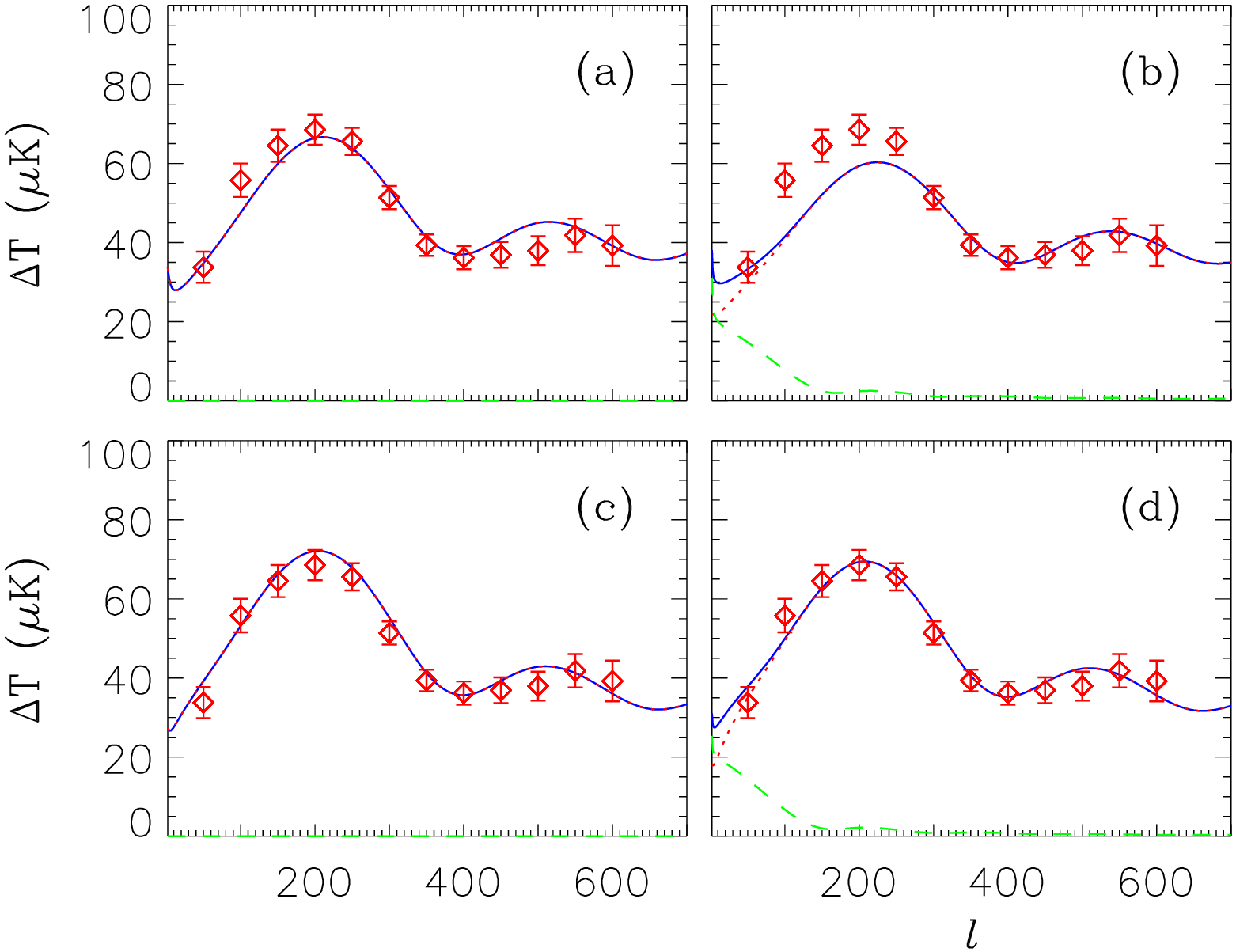,height=8.5cm,width=10cm}
\end{center}
\vspace*{0.5cm}\hspace*{1cm}
\caption{Power spectra for four different models. Panel (a) is the best
fit model with $\dlnk=0,r=0$, (b) the best fit with $\dlnk=0,r=2$, (c)
the best fit with $\dlnk \neq 0,r=0$, and (d) the best fit with $\dlnk
\neq 0,r=2$. The data points are from the Boomerang experiment
\protect\cite{boom}. The curves show: The scalar component (dotted lines),
tensor component (dashed lines), and the combined fluctuation spectrum
(solid lines).}
\label{fig5}
\end{figure}

In order to understand why allowing $\dlnk$ to vary erases any ability
to constrain $r$, it is instructive to look at the power spectra for
some of the best fit models directly.
For the remainder of this section we will assume that 
$\Omega_b h^2 = 0.019$, in accordance with Big Bang nucleosynthesis.
In Fig.~5 we show four different
power spectra, all calculated for the case of $\Omega_b h^2=0.019$.
Panels (a) and (b) are both for $\dlnk=0$. Model (a) is the best fit
with $r=0$, whereas (b) is the best fit with $r=2$.  Model (b) is a
very poor fit because introducing a tensor component while still
fitting the high $l$-values severely underestimates power around the
first peak. This cannot be remedied by shifting $n$ alone.  Therefore,
the allowed values of $r$ are quite tightly constrained.

Panels (c) and (d), on the other hand, show models where $\dlnk$ is
allowed to vary. Model (c) is the best fit for $r=0$ and (d) is the
best fit for $r = 2$. In contrast to the case with $\dlnk=0$, it is
possible to obtain a decent fit, even with a large tensor component
because the power spectrum can be ``bent'' by having a non-zero
$\dlnk$.

\begin{figure}[htb]
\begin{center}
\epsfig{file=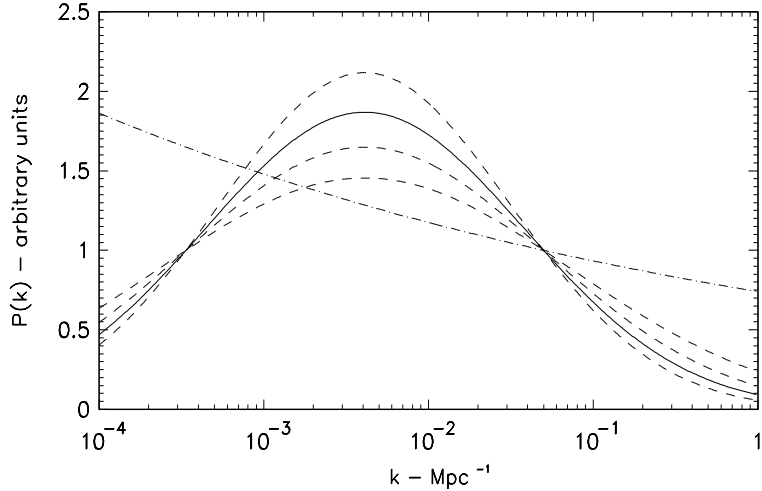,height=6.5cm,width=10cm}
\end{center}
\vspace*{-0.5cm}\hspace*{1cm}
\caption{Power spectra of scalar 
perturbations for selected values of the parameters $(n,\dlnk)$ which
provide good fits to the CMB data.  The solid line corresponds to the
best-fit point, $(n,\dlnk)=(0.5,-0.2)$.  The dashed lines correspond,
from top to bottom in the maximum region, to $(n,\dlnk)$ equal to
$(0.4,-0.24)$, $(0.6,-0.16)$, $(0.7,-0.12)$, respectively. The dot-dashed 
line corresponds to best-fit model with constant spectral index, 
i.e. $(n,\dlnk)=(0.9,0)$. All
normalizations are arbitrary.}
\label{fig6}
\end{figure}

The effect of a running spectral index can also be understood by
looking at Fig.~6, where we show the power spectra of scalar
perturbations corresponding to values of $(n,\dlnk)$ which provide
good fits to the CMB data.  The effects of a $\dlnk\neq 0$ is somewhat similar
to introduce a feature in the power spectrum at a fixed scale.
Specifically, for $\dlnk<0$, the power spectra show a maximum at a
scale $k_{m}$ given by 
\footnote{One can easily show that the
power spectrum (\ref{power}) can be rewritten as a Gaussian in $\ln k$,
centered around $k_{m}=k_{0}\exp{[-(n-1)/\dlnk]}$ and having a width 
equal to $\dlnk^{-1/2}$, i.e.:
\begin{equation}
\nonumber
{\rm ln} {\cal P} (k) = {\rm ln} {\cal P} (k_{m}) +\left.
\frac{\dlnk}{2} \right. {\rm ln} ^2 \frac{k}{k_m} + \cdots
\end{equation}}
\begin{equation}
\ln(k_{m}/k_{0})=-(n-1)/\dlnk.
\end{equation}
The CMB data provide a constraint for the possible positions of this
maximum. This can be understood by looking at the left $(n,\dlnk)$ panel
of Fig.~4, from which it is evident that the allowed region in the
plane $(n,\dlnk)$ lies along a line corresponding to $k_{m}= const$.
Specifically, the best fit model corresponds to $k_{m}=0.004 \,
\rm{Mpc}^{-1}$, while possible values are $k_{m}=0.0015 - 0.01 \,
\rm{Mpc}^{-1}$.

It is easy to understand why such a feature helps to fit the CMB data.
In order to increase sizeably the first to second peak ratio in the
CMB, one needs a power spectrum with a large negative slope at
intermediate and small scales, say e.g. $k>0.01 \;\rm{Mpc}^{-1}$.  If
we assume $\dlnk=0$ this is not allowed; the power spectrum is then a
monotonic function of $k$ and, as consequence, one automatically
obtains an excess of power at large scales, corresponding to COBE
normalization.  If instead one has $\dlnk \le 0$, the power spectrum
is non-monotonic and has a maximum at a specific scale $k_{m}$.
Specifically, if $k_m \simeq 0.004 \;\rm{Mpc}^{-1}$, the power spectrum
decreases both at large and at small scales so that one has no trouble
in reproducing the COBE data.  Clearly, as a by-product, one looses
the ability to constrain the tensor to scalar ratio $r$.  The scalar
power spectrum at large scale can in fact be strongly suppressed and
this allows for a large contribution from tensor perturbations.

We conclude this section by comparing our result with the result of
\cite{griffiths}, in which a Gaussian bump in $\log k$ was added to a
standard power spectrum and its position $k_{b}$ was constrained by
CMB data.  In very nice agreement with our results they found a best
fit $k_b$ of roughly $0.005 h \;\rm{Mpc}^{-1}$, and an allowed region
for $k_b$ of $0.001 h -0.01 h \;\rm{Mpc}^{-1}$.

\section{Conclusion}
We have considered the ability of the present day CMBR data to 
distinguish between various slow-roll models of inflation, allowing 
the scalar spectral index, $n(k)$, to vary with scale. 
Specifically, we have expanded the power spectrum ${\cal P}(k)$
to second order in $\ln k$ and we have derived the constraints
on the parameter space $(n,r,\dlnk)$ which can be obtained by using COBE,
Boomerang and MAXIMA data.

We have seen that:

\noindent
i) If we allow $\dlnk \neq 0$ the tensor to scalar ratio $r$ is
essentially unconstrained, even if we assume $\Omega_b h^2$ as low
as suggested by BBN considerations.

Moreover, assuming a BBN prior of $\Omega_b h^2= 0.019$,
we have found that:

\noindent
ii) A negative bend of the power spectrum, $\dlnk \le 0$, is favoured 
by the CMB data. 

\noindent
iii) The best fit model, $(n,\dlnk)=(0.5,-0.2)$, which provides a very
good fit to the CMB, corresponds to a power spectrum which
deviates quite strongly from a power law approximation.
The large values obtained for $n-1$ and $\dlnk$ are bordering to 
invalidate the slow-roll approximation, however, one should keep in mind
that changing the pivot scale, $k_0$, will change the value of $n-1$,
and inclusion of the 3$^{\rm rd}$ derivative of the inflaton potential
will change the value of $\dlnk$;

\noindent
iv) The CMB data favor power spectra with a bump-like feature
at scales $k_{m}=0.0015-0.01\,{\rm Mpc}^{-1}$, 
the best fit value being $k_{m}=0.004\,{\rm Mpc}^{-1}$.

Summarizing, the general result of our analysis is
that a single power law for ${\cal P}(k)$ does not provide 
a good fit to CMB data, if we assume a BBN prior $\Omega_{b} h^2=0.019$. 
In particular, CMB data favor models in which ${\cal P}(k)$ 
deviates strongly from a power law approximation. 
Specifically, the scale at which the power law approximation 
should be broken, i.e. second (or higher) order terms in eq. (\ref{power})
become important, is $k=0.0015-0.01 \,\rm{Mpc}^{-1}$.

\section*{Acknowledgements}

We wish to thank A.~Dolgov for suggestions and comments.
SHH is a Marie Curie Fellow.

\clearpage

\appendix
\section{Notation}
We use the notation: 
\[
\epsilon= \frac{M^2}{2} \left( \frac{V'}{V} \right)^2 
\, \, \, \mbox{and} \, \, \, 
\eta = M^2 \frac{V''}{V} - \frac{M^2}{2} \left( \frac{V'}{V} \right)^2~,
\]
see \cite{Kinney:2000nc} for details. 
The notation with $\eta = M^2 V''/V$,
used e.g. in \cite{Lyth:1999xn}, simply corresponds to the substitution 
$\alpha \rightarrow\alpha + 1$ in eq.(\ref{alpha_eq}). Independently on
the chosen definition for $\eta$, one finds
\be
\dlnk = -2 \xi^2 
+4 \, \frac{r}{6.8} \left[ (n-1) + \frac{3}{2} \frac{r}{6.8} \right]~,
\ee
where $\xi^2 \equiv M^4 V'V'''/V^2$.
One could potentially include higher order terms in Eq.~(1),
corresponding to $d^2n/d {\rm ln} k^2 \neq 0$, which would be
expressed as
\begin{eqnarray}
\frac{d^2n_S}{d {\rm ln}k^2} & = & 2 \sigma^3 + 
\frac{1}{2} \dlnk \left( 9 \frac{r}{6.8} - (n-1) \right) \\
&&- 2 \left( n-1\right)^2 \frac{r}{6.8} 
- 15 \left( n-1\right)\left( \frac{r}{6.8}\right)^2 -
15\left( \frac{r}{6.8}\right)^3~,
\\
\frac{d^2n_T}{d {\rm ln}k^2} & = & \dlnk \frac{r}{6.8} -
\left( n-1\right)^2 \frac{r}{6.8} -
3 \left( n-1\right) \left(  \frac{r}{6.8} \right)^2   - 
2\left( \frac{r}{6.8}\right)^3~,
\end{eqnarray}
with $\sigma^3 \equiv 2 \epsilon V''''/V$.

\end{document}